\documentclass[a4paper, textwidth = 170mm]{extarticle}

\usepackage{graphicx}
\usepackage{epsfig}

\begin{document}

\title{Planetary motion on an expanding locally anisotropic background}

\author{P. Castelo Ferreira}




\date{}
\maketitle

\begin{abstract}
In this work are computed analytical solutions for orbital motion on a background described by an Expanding Locally Anisotropic (ELA) metric ansatz. This metric interpolates between the Schwarzschild metric near the central mass and the Robertson-Walker metric describing the expanding cosmological background far from the central mass allowing for a fine-tuneable covariant parameterization of gravitational interactions corrections in between these two asymptotic limits. In particular it is shown that the decrease of the Sun's mass by radiation emission plus the General Relativity corrections due to the ELA metric background with respect to Schwarzschild backgrounds can be mapped to the reported yearly variation of the gravitational constant $\dot{G}$ through Kepler's third law. Based on the value of the heuristic fit corresponding to the more recent heliocentric ephemerides of the Solar System are derived bounds for the value of a constant parameter $\alpha_0$ for the ELA metric as well as the maximal corrections to perihelion advance and orbital radii variation within this framework. Hence it is shown that employing the ELA metric as a functional covariant parameterization to model gravitational interactions corrections within the Solar System allows to maintain the measurement projection standards constant over time, specifically both the Astronomical Unit ($AU$) and the Gravitational constant ($G$). Also it is noted that the effect obtained is not homogeneous for all planetary orbits consistently with the diversity of estimates in the literature obtained assuming Schwarzschild backgrounds.
\end{abstract}

\section{Introduction\label{sec.introd}}

\subsection{Motivation}

In this work our main objective is to employ a background described by an expanding locally anisotropic (ELA) metric~\cite{PLB,e_print} to analytically model the corrections to orbital motion with respect to orbital motion on a Schwarzschild background within the Solar System. In particular, based in analytical orbital solutions for the background described by the ELA metric we will explicitly compute the respective corrections to the orbital period due to two distinct contributions, both the Sun's mass decrease corrections and the corrections obtained when considering expanding backgrounds~\cite{review}. In particular it is shown that, when considering a expanding background described by the ELA metric, it is possible to map these corrections to the heuristic variation of the gravitational constant $\dot{G}$, simultaneously having negligible contributions to the remaining orbital parameters. Considering the most recent estimate for the average value of the variation $\dot{GM}_\odot=\dot{G}\times M_\odot+G\times\dot{M}_\odot$ obtained from the numerical analysis of planetary ephemerides~\cite{dG} we will compute bounds for the values of the metric functional parameter which, in turn, will allow to estimate the contributions due to the ELA metric background to the orbital period, orbital perihelion advance and orbital radius variation of the several planets within the Solar System.

The ansatz for the Expanding Locally Anisotropic (ELA) metric~\cite{PLB} was originally suggested as a possible description of local matter distributions in the expanding universe~\cite{Hubble,Inflation}, hence interpolating between the Schwarzschild (SC) metric~\cite{Schwarzschild} near the central mass and the Robertson-Walker (RW) metric~\cite{RW} describing the expanding cosmological background far from the central mass. Hence the ELA metric generalizes the isotropic McVittie metric~\cite{McVittie} and the anisotropic metrics considered in~\cite{SCS}, having the novelty of maintaining the SC event horizon free of space-time singularities and maintain as the only space-time singularity the SC mass pole at the origin such that the value of the SC mass pole is maintained~\cite{sing,PLB}. This metric is locally anisotropic consistently with astrophysical observations~\cite{anisotropy,WMAP} converging at large radius to the RW metric such that global isotropy is maintained. As this metric depends on a functional parameter $\alpha(r)$ dependent on the radial distance $r$ to the central mass being considered, only the asymptotic limits for this parameter are fixed, specifically at spatial infinity, at the SC radius (the event horizon) and at the location of massive point-like particles. In between the SC radius and spatial infinity there are no theoretical restrictions on the metric parameter $\alpha$.

We recall that the two asymptotic backgrounds for the ELA metric (the SC metric and the RW metric) are well established results both phenomenologically and theoretically such that an interpolation between both these limiting solutions is physically required justifying the existence of gravitational corrections to SC backgrounds at intermediate spatial scales. In addition let us note that the background mass-energy density as well as the corrections to the gravitational acceleration with respect to the respective quantities for SC backgrounds depend on the metric functional parameter $\alpha$ such that the value of this parameter allows to fine-tune the gravitational interactions corrections, namely the predictions for orbital motion and gravitational red-shifts. Therefore the motivation to investigate whether measured deviations from the predictions on SC gravitational backgrounds can be consistently described by an ELA background is two-fold:
\begin{itemize}
\item it allows for the definition of the functional parameter $\alpha$ for intermediate spatial scales maintaining compatibility with the asymptotic solutions, hence allowing to research the inclusion of the ELA metric within more fundamental frameworks;
\item it allows for a mathematical modelation of gravitational systems maintaining compatibility with the asymptotic well established solutions, hence simultaneously increasing the accuracy of predictions and fixing a relation between the corrections to gravitational acceleration, red-shifts and the background mass-energy density through Einstein Equations.
\end{itemize}
Hence, generally, a fit of the gravitational parameter to experimental data allows both for a modelation parameterization of gravitational corrections to SC backgrounds as well as to fix, within this framework, the background mass-energy density such that this construction
describes unaccounted background extended matter densities as well as the respective gravitational corrections to physical observables similarly to the heuristic dark matter distributions~\cite{curves}.

\subsection{ELA Metric}
Specifically the line-element for the ELA metric is~\cite{PLB}
\begin{equation}
\begin{array}{rcl}
ds^2&=&\displaystyle\left(1-U_{\mathrm{SC}}\right)c^2\,dt^2-r^2\left(d\theta^2+\sin^2\theta d\varphi^2\right)\\[6mm]
& &\displaystyle-\frac{1}{1-U_{\mathrm{SC}}}\left(dr-H\frac{\,r}{c}\left(1-U_{\mathrm{SC}}\right)^{\frac{\alpha}{2}+\frac{1}{2}}c\,dt\right)^2\ .
\end{array}
\label{g_generic}
\end{equation}
where $H=\dot{a}/a$ is the time dependent Hubble rate defined as the rate of change of the universe scale factor $a$, $U_{\mathrm{SC}}=2GM/(c^2r)$ is the usual Schwarzschild gravitational potential, $G$ is the Gravitational constant, $M$ is the value of the Schwarzschild mass pole for the central mass being considered and $c$ is the speed of light in vacuum. Here the radial coordinate $r$ cohincides with the physical measurable distances, hence corresponds to the proper spatial coordinate. The exponent $\alpha$ is, generally, a function of the radial coordinate which must obey the following asymptotic conditions~\cite{PLB}:
\begin{enumerate}
\item $\displaystyle\alpha(r\sim 0)\sim -\frac{1}{r}$, ensuring that the value of the SC mass pole is maintained;
\item $\alpha(r=r_{\mathrm{SC}})\geq 3$, ensuring that the SC event horizon is not an extended space-time singularity;
\item $\alpha(r\sim +\infty)\sim 0$, ensuring that the total background mass is finite and the RW background is asymptotically recovered;
\end{enumerate}
where $r_{\mathrm{SC}}=2GM/c^2$ is the SC radius. Following these conditions we consider the simplified ansatz studied in~\cite{PLB,Pioneer_ELA}
\begin{equation}
\begin{array}{rcl}
\displaystyle\alpha(r)&=&\displaystyle(\bar{\alpha}_0-\alpha_1)+\alpha_1\,U_{\mathrm{SC}}(r)=(\bar{\alpha}_0-\alpha_1)+\alpha_1\,\frac{2GM}{c^2\,r}\ ,\\[5mm]
\displaystyle\bar{\alpha}_0&=&\displaystyle\left\{\begin{array}{lcl}\displaystyle \alpha_{0.0}\geq 3&,&\displaystyle r\sim r_{\mathrm SC}\\[5mm] \displaystyle \alpha_0&,&\displaystyle r\gg r_{\mathrm SC}\end{array}\right.
\end{array}
\label{alpha_r}
\end{equation}
where $\bar{\alpha}_0$ and $\alpha_1$ are numerical coefficients. The bound $\bar{\alpha}_0\ge 3$ ensures that the SC radius  is an event horizon and space-time is singularity free at this horizon, while for the bound $\bar{\alpha}_0>5$ space-time is asymptotically Ricci flat near the event horizon such that the SC metric is a good approximation in a neighbourhood of the point-like mass $M$. The coefficient $\alpha_1<0$ ensures that the singularity at the origin coincides with the SC mass-pole. In the following analysis we will consider it arbitrarily close to zero, $-1\ll \alpha_1< 0$, such that outside the event horizon its effects are negligible, hence $\alpha\approx \alpha_0$ for planetary orbits. For further details in the derivation of the ELA metric ansatz see~\cite{PLB}.

Based in this ansatz we will compute the analytical corrections to two body heliocentric orbital motion within the solar system on the ELA metric background~(\ref{g_generic}) with respect to orbital motion on Schwarzschild backgrounds~\cite{Kenyon,Gravitation}. More accurate results including the many body gravitational interactions in the Solar System can only be computed by extensive numerical analysis including the known bodies in the solar system~\cite{Pitjeva,Fienga}. These analysis are usually carried in the PPN formalism which includes both the General Relativity corrections to the classical Newton law of gravitation on Schwarzschild backgrounds, as well as corrections of extended theories of gravity such as Brans-Dicke gravity~\cite{PPN}.

\subsection{Standard of measurement in the Solar System}
In addition to the IS unit (the meter), as standard of measurement within the Solar System it is also commonly employed the Astronomical Unit ($AU$). Today this quantity has a fixed value in IS units. However its original definition relies on the classical Kepler's third law fixing a mathematical relation between the measurement projection for spatial-lengths and the measurement projection for temporal-lengths (through the gravitational constant $G$). Specifically the original definition of the $AU$ was stated in IS units as~\cite{AU_def,AU_1,AU_2,dG}
\begin{equation}
AU=\left(G\,M_\odot\,\left(\frac{T_{1AU.0}}{2\pi}\right)^2\right)^\frac{1}{3}=149597870700\,m\ ,
\label{AU}
\end{equation}
where $M_{\odot}=1.9891\times 10^{30}\,kg$ is the Sun's mass, $G=6.67\times 10^{-11}\,kg^{-1}\,m^3\,s^{-2}$
is the Gravitational constant and $T_{1AU.0}=31562889.928\ldots\,s$ is the Keplerian orbital period for a point mass in an elliptic orbit with semi-major axis of value $1\,AU$. This original definition exactly matches Kepler's third law for a planet orbiting the Sun in an elliptic orbit with semi-major axis of length $r_{\mathrm{orb}}=1\,AU$ being based on the existence of a classical Keplerian constant of motion, specifically the angular momentum $J_0(r_{\mathrm{orb}}=1\,AU)=-\sqrt{G\,M\,AU\,(1-e^2)}$, where $e$ is the orbit eccentricity. Hence time and space measurements are related by this classical conservation law stated in the definition~(\ref{AU}).

As many of simulations and analysis of dynamical motion within the Solar System, in particular of orbital motion, in the literature assumed
this constraint~(\ref{AU}), often the interpretation of experimental data would favour either a time varying value for the $AU$ or the gravitational constant $G$ as the best fits. In particular, in the original work by Krasinsky and Brumberg~\cite{AU_1} (see also~\cite{AU_2}), considering experimental data from range measurements of planetary orbital motion plus range measurements from orbiters and landers it was concluded that a yearly variation of the $AU$ constitutes a best fit to the planetary and spacecraft ephemerides within the Solar System.
It is today accepted that such variation has no physical meaning and that any standard of measurement should be maintained fixed~\cite{dG,Fienga}.
Nevertheless these analysis imply that some sort of unmodeled gravitational interaction deviation with respect to Schwarzschild backgrounds
predictions is present in the Solar System.

We also note that formally, any corrections to Kepler's third law~(\ref{AU}), can be mathematical rewritten as a time varying $AU$ or $G$ such that maintaining the classical constraint~(\ref{AU}) for a fixed value of the $AU$ accounts for a time varying $G$. Following these observations the planetary ephemerides were recently re-analyzed by Pitjeva and Pitjev~\cite{dG} maintaining the $AU$ fixed and accounting for the experimental bounds on the Sun's mass variation $\dot{M}_\odot\neq 0$ showing that, further including a time varying Gravitational constant $\dot{G}\neq 0$, allows for a better fit than previous analysis. Specifically the estimate considered for the Sun's mass variation due to radiation and matter emission and absorption is (see~\cite{dG,Uzan} and references therein)
\begin{equation}
\frac{\dot{M}_\odot}{M_\odot}=-6.7^{+3.1}_{-3.1}\times 10^{-14}\,(yr^{-1})\ ,
\label{dAU_dM}
\end{equation}
and the unmodeled fitted value of the variation of the Gravitational constant is~\cite{dG}
\begin{equation}
\frac{\dot{G}}{G}=+1.65^{+5.85}_{-5.85}\times 10^{-14}\,(yr^{-1})\ .
\label{dAU_dG}
\end{equation}

In the remaining of this work we exploit the mathematical constraint between the Gravitational constant and the orbital period implemented by Kepler's third law, hence the original definition of the $AU$~(\ref{AU}), showing that the perturbative corrections to the classical Keplerian period due to the background described by the ELA metric~(\ref{g_generic}) can describe this unmodeled variation of $G$. Generally such map is formally expressed by considering the corrections to the several quantities over a given period of time, for instance an orbital period for a orbit with semimajor axis of value $1\,AU$. For the specific case of the $AU$ definition~(\ref{AU}), maintaining a fixed value for this standard of measurement, we obtain
\begin{equation}
\begin{array}{rcl}
\displaystyle\frac{\dot{G}}{G}&\equiv&\displaystyle\frac{1}{T_{1AU.0}}\frac{\Delta G_{1AU.0}}{G}\\[5mm]
&=&\displaystyle-\frac{1}{T_{1AU.0}}\left(\frac{\Delta M_{\odot.1AU.0}}{M_\odot}+2 \frac{AU^3}{GM_\odot}\left(\frac{2\pi}{T_{1AU.0}}\right)^2\frac{\Delta T_{1AU.0}}{T_{1AU.0}}\right)\ ,
\end{array}
\end{equation}
where $\Delta G_{1AU.0}$ is the unmodeled variation of the Gravitational constant $G$ over one orbital period, $\Delta M_{\odot.1AU.0}$ is the
variation of Sun's mass due to emission and absorption of matter and radiation over one orbital period and $\Delta T_{1AU.0}$ is the modeled variation (over one orbital period) of the orbital period with respect to the classical Keplerian orbital period $T_{1AU.0}$. We remark that for a
given planetary orbit, also the semimajor axis will generally be varying over time.

We also note that the estimates of~\cite{Pitjeva,dG} are computed for heliocentric distances and the estimates of~\cite{Fienga} are computed
for barycentric distances such that some discrepancies, mainly on estimates for orbital radii and time variation of the orbital radii are verified between both approaches as the distance between the centre of the Sun and the solar System barycentre is of order $\sim 10^8\,m$. For technical simplification and aiming at a direct comparison with the estimates in~\cite{Pitjeva,dG}, we will carry our calculations in heliocentric distances. We stress that for a numerical simulation on an expanding background with a fix value of the $AU$ and without considering the classical constraint corresponding to the Kepler third law~(\ref{AU}) the gravitational constant should be considered fixed such that the deviation of orbital motion from the predictions of Schwarzschild backgrounds is fully described by the corrections to gravitational interactions which depend on the functional parameter $\alpha(r)$~(\ref{g_generic}).

\subsection{Outline}
When required, for numerical evaluation of the Hubble rate $H$ and the deceleration factor $q$ of today's universe, we are considering the values~\cite{WMAP}
\begin{equation}
\begin{array}{rcl}
H_0&=&\displaystyle\left.H\right|_{t=t_0}=2.28\times 10^{-18}\,s^{-1}\ ,\\[5mm] q_0&=&\displaystyle\left.-\frac{\ddot{a}}{a}\left(\frac{\dot{a}}{a}\right)^{-2}\right|_{t=t_0}=-0.582\ .
\end{array}
\end{equation}
As for the planetary orbital parameters considered for numerical evaluations we are considering the data presented in table~\ref{table.planet_data}.
\begin{table}[ht]
\begin{center}
{\small
\begin{tabular}{lcccccc}
\hline\noalign{\smallskip}
Planet  &$r_{\mathrm{orb}}$&$e$ &$m$&$\mathrm{\delta_{orb}}$&$\mathrm{\delta_{per}}$&$f$\\
 &$(\times 10^{11}\,m)$& &$(\times 10^{24}\,kg)$&$\mathrm{(deg)}$&$\mathrm{(deg)}$&$\mathrm{(deg)}$\\
\noalign{\smallskip}\hline\noalign{\smallskip}
Mercury &$0.579091768$&$0.20563069$&$0.3302$ &$7.00487$ &$77.45645$&$174.796$\\
Venus   &$1.08208926$ &$0.00677323$&$4.8685$ &$3.39471$ &$131.53298$&$50.4468$\\
Earth   &$1.49597887$ &$0.01671022$&$5.9736$ &$0$       &$102.94719$&$357.517$\\
Mars    &$2.27936637$ &$0.09341233$&$0.64185$&$1.85061$ &$336.04084$&$19.3564$\\
Jupiter &$7.78412027$ &$0.04839266$&$1898.6$ &$1.30530$ &$14.75385$&$18.8180$\\
Saturn  &$14.26725413$&$0.05415060$&$568.46$ &$2.48446$ &$92.43194$&$320.347$\\
Uranus  &$28.70972220$&$0.04716771$&$86.832$ &$0.76986$ &$170.96424$&$142.956$\\
Neptune &$44.98252911$&$0.00858587$&$102.43$ &$1.76917$ &$44.97135$&$267.767$\\
Pluto   &$59.06376272$&$0.24880766$&$0.0125$ &$17.14175$&$224.06676$&$14.8601$\\
\noalign{\smallskip}\hline
\end{tabular}}
\end{center}
\caption{Planetary orbits parameters: the semi-major axis
	$r_{\mathrm{orb}}$, the eccentricity $e$, the mass $m$, the orbital inclination $\delta_{orb}$, the longitude of perihelion $\delta_{per}$ and the true anomaly $f$~\cite{NASA}.}
\label{table.planet_data}
\end{table}

This work is organized as follows. In section~\ref{sec.orbits} are computed the analytical solutions for orbital motion on the background describe by the ELA metric being derived the General Relativity corrections to orbital perihelion advance and orbital period for such backgrounds. This derivation is carried considering a static elliptical orbit approximation. In section~\ref{sec.radius} are analyzed circular orbits on the ELA metric background~(\ref{g_generic}) and computed the orbital radius variation within this approximation. In section~\ref{sec.AU} are analyzed the corrections to Kepler's third law due to the decrease of the Sun's mass and due to the background described by the ELA metric. In particular are derived estimates for the several contributions that allow to match the heuristic fit to the variation of the Gravitational constant~\cite{dG} and it is computed the value of the ELA metric parameter $\alpha_0$ that maps such fit to the gravitational corrections obtained for the background described by the ELA metric. Are also computed the respective corrections to orbital perihelion advance and orbital radius variation. In the conclusions we shortly resume and discuss the results obtained in this work.

\section{Perturbative Static Elliptical Orbit Solutions\label{sec.orbits}}

In this section we derive analytical solutions for elliptical orbits on the background described
by the ELA metric ansatz~(\ref{g_generic}). In particular we will explicitly compute the General
Relativity corrections on such backgrounds to the Keplerian elliptical orbit solution $r(\varphi)=1/u_0(\varphi)$ with
\begin{equation}
u_0(\varphi)=\frac{1+e\,\cos(\varphi)}{d}\ \ ,\ \ d=r_{\mathrm{orb}}(1-e^2)\ ,
\label{u0.0}
\end{equation}
where $e$ is the orbit eccentricity and $r_{\mathrm{orb}}$ is the elliptic orbit semi-major axis. We note
that such corrections will generally include the General Relativity corrections on Schwarzschild
backgrounds as well as corrections depending on the Hubble rate $H$.

It is hard, if not impossible, to obtain an exact analytical solution considering the differential equations for a time varying Hubble rate $H$. The main difficulty is that energy conservation is no-longer given by a constant of motion, instead we have a non-linear second order differential equation on the function $t$ coupled to the differential equation for $r$. Hence, for technical simplification purposes, we are taking the static orbit approximation by considering a fixed Hubble rate $H=H_0$ corresponding to today's measured value for this rate.

From the metric line-element~(\ref{g_generic}) let us consider the Lagrangian definition~\cite{Kenyon,Gravitation} to order $H_0^2$
\begin{equation}
\begin{array}{rcl}
\displaystyle\frac{{\mathcal{L}}}{m}&=&\displaystyle\left(1-U_{\mathrm{SC}}-\left(H_0\,\frac{r}{c}\right)^2\left(1-U_{\mathrm{SC}}\right)^\alpha\right)\,\left(c\,\frac{dt}{d\tau}\right)^2\\[6mm]
&&\displaystyle+2H_0\,\frac{r}{c}\,\left(1-U_{\mathrm{SC}}\right)^{\frac{\alpha_0}{2}-\frac{1}{2}}\,c\,\frac{dt}{d\tau}\,\frac{dr}{d\tau}-\frac{1}{1-U_{\mathrm{SC}}}\left(\frac{dr}{d\tau}\right)^2-r^2\left(\frac{d\varphi}{d\tau}\right)^2\\[6mm]
&&\displaystyle+O\left(H_0^2t\right)\ .
\end{array}
\label{A.L_orbit}
\end{equation}
This Lagrangian is a constant ${\mathcal{L}}/m=c$ and it is considered that the orbit of the test body is lying in the plane of constant coordinate $\theta=\pi/2$ such that $d\theta=0$ and $\sin\theta=1$~\cite{Kenyon,Gravitation}. The Lagrangian is independent of the coordinate $\varphi$, hence
a constant of motion corresponding to angular momentum exists being given by the variational
derivation of the Lagrangian with respect to $d\varphi/d\tau$,
\begin{equation}
J=\frac{1}{2m}\frac{\delta{\mathcal{L}}}{\delta\frac{d\varphi}{d\tau}}=-r^2\,\frac{d\varphi}{d\tau}\ .
\label{J}
\end{equation}
Also, due to the Lagrangian~(\ref{A.L_orbit}) not depending explicitly on the time coordinate, a conserved constant of motion corresponding
to energy exists being given by the functional variation of the Lagrangian with respect to $c\,dt/d\tau$
\begin{equation}
\begin{array}{rcl}
\displaystyle\frac{2E_H}{m\,c}&=&\displaystyle\frac{1}{m}\,\frac{\delta{\mathcal{L}}}{\delta(c\,dt/d\tau)}\\[6mm]
&=&\displaystyle 2\left(1-U_{\mathrm{SC}}-\left(H_0\,\frac{r}{c}\right)^2\left(1-U_{\mathrm{SC}}\right)^\alpha\right)\,\left(c\,\frac{dt}{d\tau}\right)\\[6mm]
&&\displaystyle+2H_0\,\frac{r}{c}\,\left(1-U_{\mathrm{SC}}\right)^{\frac{\alpha_0}{2}-\frac{1}{2}}\,\left(\frac{dr}{d\tau}\right)\ .
\end{array}
\end{equation}
This equation can be solved for $c\,dt/d\tau$ such that replacing the obtained solution in the Lagrangian~(\ref{A.L_orbit}), expressing
the derivatives with respect to proper time $dr/d\tau$ by the derivatives with respect to $\varphi$,
$dr/d\tau=dr/d\varphi\times d\varphi/d\tau$ and considering the change of variables $u=1/r$,
further differentiating with respect to $\varphi$ and factoring out an overall factor of $2u'J^2$ (with the primed quantities representing derivation
with respect to $\varphi$), we obtain the approximate differential equation of order $H_0^2$ for the function $u(\varphi)$ describing an orbiting test mass in the gravitational field of a point-like central mass $M$
\begin{equation}
\begin{array}{rcl}
u''+u&=&\displaystyle\frac{GM}{J^2}+\frac{3GM}{c^2}\,u^2\\[6mm]
&&\displaystyle -\alpha_0\,\frac{GM}{c^2}\left(\frac{H_0}{c}\right)^2\,\left(1-\frac{2GM}{c^2}\,u\right)^{-1+\alpha_0}\\[6mm]
&&\displaystyle -\left(\frac{H_0}{J}\right)^2\,\frac{1}{u^3}\,\left(1-\frac{2GM}{c^2}\,u\right)^{-1+\alpha_0}\,\left(1-\frac{2GM}{c^2}\,u+\frac{\alpha_0\,GM}{c^2}\,u\right)\ .\\[6mm]
\end{array}
\label{A.Eq_u}
\end{equation}
The terms in the first line match the usual terms obtained for Schwarzschild backgrounds and the terms in the second and third lines are the corrections due to the ELA metric background.

We note that, although maintaining the terms of order $H_0^2$ in the Lagrangian~(\ref{A.L_orbit}) which do not depend explicitly on the time coordinate we have neglected one term containing the factor $H_0^2\,t$. Explicitly it is the term $-2q_0H_0^2\,t\,r\,(1-U_{\mathrm{SC}})^{(\alpha_0-1)/2}(dt/d\tau)\,(dr/d\tau)$.
Comparing the terms of order $H_0^2$ in the Lagrangian with this term we conclude that this is a valid approximation as long as the value of the time coordinate is below the following bound
\begin{equation}
t\ll -\frac{1}{q_0}\,\frac{r}{dr/d\tau}\,(1-U_{\mathrm{SC}})^{\frac{\alpha_0}{2}-\frac{1}{2}}\,\frac{E_H}{mc^2}\sim 10^{13}\ years.
\label{t_bound}
\end{equation}
This bound is well above any astrophysical measurement time span and has been obtained by considering the following simplified assumptions,
 within the solar system, from the experimental upper bounds on the orbital radius variations within the Solar System~\cite{Uzan} we consider
the estimate for the ratio $|r/\dot{r}_1|> 10^{20}$, assume weak gravitational field $U_{\mathrm{SC}}\ll 1$ and values of $\alpha_0$ for which the approximation $(1-U_{\mathrm{SC}})^{(\alpha_0-1)/2}\sim 1$ is valid. We note that this approximation will no longer be valid for very large values of the metric exponent, $|\alpha_0|\gg 0$.

Noting that for orbits in the solar system the function $u$ has relatively small values ($0.5\times 10^{-12}<u<0.5\times 10^{-10}\,m^{-1}$, where $r_{\mathrm{orb}}$ is the orbit semi-major axis), with the objective of further simplifying the differential equation~(\ref{A.Eq_u}), we consider a series expansion on $u$ of the terms of order $H_0^2$ which is equivalent to an expansion on the weak gravitational field. Specifically, for a generic exponent $p$, the factor $(1-U_{\mathrm{SC}})^p$ has the following series expansion
\begin{equation}
\begin{array}{rcl}
\displaystyle\left(1-\frac{2GM\,u}{c^2}\right)^p&=&\displaystyle 1-p\,\frac{2GM}{c^2}\,u+p\left(p-1\right)\left(\frac{2GM}{c^2}\right)^2\,\frac{u^2}{2}\\[2mm]
&&\displaystyle-p\,(p-1)(p-2)\left(\frac{2GM}{c^2}\right)^3\,\frac{u^3}{6}+O\left(p^4\left(\frac{2GM}{c^2}\right)^4\,u^4\right) \ .
\end{array}
\end{equation}
We note that the full series is strictly convergent independently of the value of the exponent $p$ as long as $u<c^2/(2GM)$, however an approximation to first order on $u$ will only be valid as long as $2p\,GM/c^2\,u<1$, otherwise it is required to consider higher order terms to attain a valid approximation. Hence the differential equation~(\ref{A.Eq_u}) is, to order $u^2$, rewritten as
\begin{equation}
\begin{array}{rcl}
u''(\varphi)+A\,u(\varphi)&\approx&\displaystyle \frac{GM}{J^2}\,B+\frac{3GM}{c^2}\,C\,u^2\\[6mm]
&&\displaystyle-\left(\frac{H_0}{J}\right)^2\,\frac{1}{u^3}+\alpha_0\left(\frac{H_0}{J}\right)^2\,\frac{GM}{c^2}\,\frac{1}{u^2}+O\left(u^3\right)\ .
\end{array}
\label{orbits_exp}
\end{equation}
We note that with respect to the General Relativity orbit's equation on Schwarzschild backgrounds, there are the extra multiplicative factors $A=1+\delta_A$, $B=1+\delta_B$ and $C=1+\delta_C$. These factors differ from unity by the following additive constants
\begin{equation}
\begin{array}{rcl}
\delta_A&=&\displaystyle-2(\alpha_0-1)\alpha_0\left(\frac{GM\,H_0}{c^3}\right)^2\left(1+(\alpha_0^2-5\alpha_0+6)\frac{(GM)^2}{3c^2\,J^2}\right)\ ,\\[6mm]
\delta_B&=&\displaystyle-\alpha_0\,\left(\frac{H_0}{c^2}\right)^2\left(J^2+(\alpha_0^2-3\alpha_0+2)\frac{2(GM)^2}{3c^2}\right)\ ,\\[6mm]
\delta_C&=&\displaystyle-\frac{2}{3}\alpha_0\left(\alpha_0^2-3\alpha_0+2\right)\,\left(\frac{GM\,H_0}{c^3}\right)^2\left(1+(\alpha_0^2-7\alpha_0+12)\frac{(GM)^2}{5c^2\,J^2}\right)\ .
\end{array}
\label{A_B_C}
\end{equation}

\begin{figure}
\includegraphics[width=\textwidth]{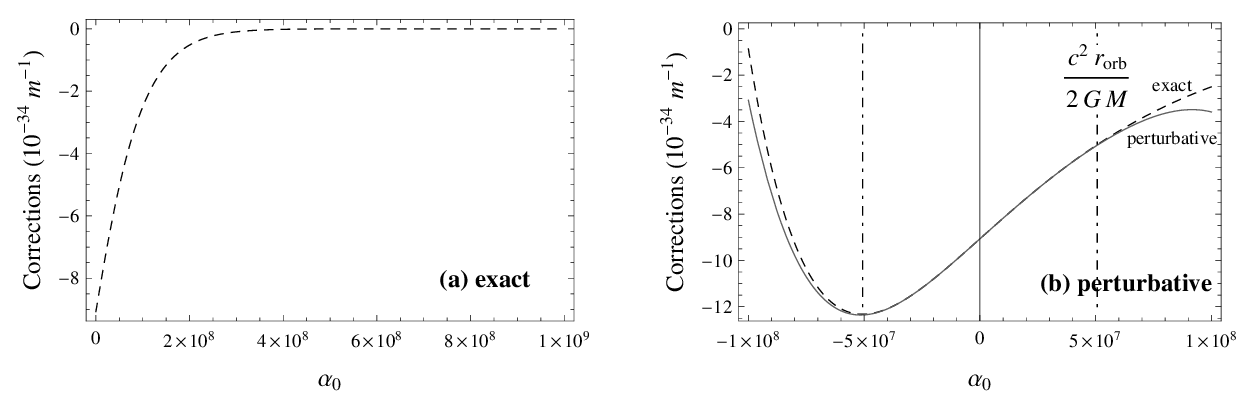}
\caption{Plot of the exact (dashed line) and perturbative
(continuous line) expressions for the corrections of the orbital differential
equation~(\ref{A.Eq_u}) as a function of the
parameter $\alpha_0$ for Earth's orbit due to the to ELA metric background with respect to Schwarzschild background. The perturbative regime is valid for $|\alpha_0|<\alpha_{0.\mathrm{max.pert}}\approx 5\times 10^7$~(\ref{alpha_max_pert}):\hfill\break
\ {\bf(a)} plot of the exact expressions for $\alpha_0 > 0$, the corrections asymptotically vanish for large $\alpha_0\gg \alpha_{0.\mathrm{max.pert}}$;\hfill\break
\ {\bf(b)} plot of the exact and perturbative expressions for $\alpha_0\in]-10^8,10^8[$, the perturbative and exact expressions approximately match up to $|\alpha_0|=\alpha_{0.\mathrm{max.pert}}$.}
\label{fig.orbits_perturbative}
\end{figure}

So far we have not specify for which values of the coefficient $\alpha_0$ the static approximation considered is valid. By comparing the leading order terms with the next to leading order terms we conclude that this perturbative equation is valid only for absolute values of the parameter $\alpha_0$ up to
\begin{equation}
|\alpha_0|\,<\,\alpha_{0.\mathrm{max.pert}}\approx \frac{c^2\,r_{\mathrm{orb}}}{2GM}\ .
\label{alpha_max_pert}
\end{equation}
Above this value it is either necessary to consider higher order terms of the series expansion or to consider the exact expressions. Nevertheless we remark that, for a fixed positive value of the radial coordinate $r$, and larger positive values of the parameter $\alpha_0>\alpha_{0.\mathrm{max.pert}}$ the corrections given by the exact expression due to the ELA metric background will decrease significantly in absolute value becoming, for very large values of the parameter $\alpha_0\gg \alpha_{0.\mathrm{max.pert}}$, negligible, while for larger negative values of the parameter $\alpha_0<-\alpha_{0.\mathrm{max.pert}}$ the corrections become more significant being unbounded from below. Hence, although for positive values of $\alpha_0$ the approximation~(\ref{orbits_exp}) subject to the bound~(\ref{alpha_max_pert}) allows to establish a fairly good estimate for the maximum contribution of the corrections on the ELA metric background, for negative values of $\alpha_0$ no bounds can be set for such contribution.  In figure~\ref{fig.orbits_perturbative} are plotted the values of the exact and perturbative correction terms of order $H_0^2$ in the differential equation~(\ref{A.Eq_u}).

In addition, with respect to the bound~(\ref{t_bound}), we note that it is also obeyed as long as the bound~(\ref{alpha_max_pert}) is obeyed, hence for larger values of the coefficient $\alpha_0$ the terms of order $H_0^2$ explicitly depending on the time coordinate become relevant and must be included in the Lagrangian~(\ref{A.L_orbit}). For these cases there is no constant of motion directly associated with energy conservation. Specifically, conserved energy, would be given by the constant $E=\int d\tau(\delta{\mathcal{L}}/\delta \dot{t}-d/d\tau(\delta{\mathcal{L}}/\delta t))$, such that, due to the complexity of the full equations of motion, it would be preferable to consider a numerical analysis to compute orbital motion.

We are proceeding assuming that the upper bound~(\ref{alpha_max_pert}) is obeyed. To solve the differential equation~(\ref{orbits_exp}) we start by solving the differential equation considering only the dominant term in the right-hand side of~(\ref{orbits_exp}), hence obtaining~\cite{Kenyon,Gravitation}
\begin{equation}
u_{0.H^2}''+A\,u_{0.H^2}=\frac{GM}{J^2}\,B\ \ \Rightarrow\ \ u_{0.H^2}=\frac{1+e\,\cos(\sqrt{A}\,\varphi)}{d}\ .
\label{u_0_H}
\end{equation}
The standard General Relativity angular momentum $J_0$ and the angular momentum $J$ are expressed in terms of the parameter $d$ as
\begin{equation}
\begin{array}{rcl}
J_0&=&\displaystyle-\sqrt{G\,M\,d}\ ,\\[5mm]
J&=&\displaystyle-\sqrt{G\,M\,d\,\frac{B}{A}}\approx \frac{1}{2}\,J_0\left(\delta_B-\delta_A\right)_{J=J_0}\ ,
\end{array}
\label{J_d_H}
\end{equation}
where to evaluate $A$ and $B$, we have approximated the angular momentum by the respective Keplerian quantity, $J\approx J_0$.

Next let us compute the corrections to the solution $u_{0.H^2}$ by considering the remaining terms in the right-hand side of the differential equation~(\ref{orbits_exp}) evaluated for the function $u_{0.H^2}$~(\ref{u_0_H}) such that the full solution is
\begin{equation}
u=u_{0.H^2}+u_{\mathrm{GR}.H^2}+u_{H^2}\ .
\label{u_H2}
\end{equation}
Here the functions $u_{\mathrm{GR}.H^2}$ and $u_{H^2}$ correspond respectively to the corrections to the Keplerian orbit's solution due to the Schwarzschild background and due to the ELA metric background approximated to order $H_0^2$ being, respectively, the solutions of the following differential equations
\begin{equation}
\begin{array}{rcl}
\displaystyle
u_{\mathrm{GR}.H^2}''+A\,u_{\mathrm{GR}.H^2}&=&\displaystyle\frac{3GM}{c^2}\,C\,u_{0.H^2}^2=\frac{3GM}{c^2}\,C\,\frac{(1+e\cos(\sqrt{A}\,\varphi))^2}{d^2}\ ,\\[6mm]
\displaystyle u_{H^2}''+A\,u_{H^2}&=&\displaystyle -\left(\frac{H_0}{J}\right)^2\,\frac{1}{u_{0.H^2}^3}+\alpha_0\left(\frac{H_0}{J}\right)^2\,\frac{GM}{c^2}\,\frac{1}{u_{0.H^2}^2}\\[6mm]
&=&\displaystyle -\left(\frac{H_0}{J}\right)^2\,\frac{1}{(1+e\cos(\sqrt{A}\,\varphi))^3}\\[6mm]
&&\displaystyle+\alpha_0\left(\frac{H_0}{J}\right)^2\,\frac{GM}{c^2}\,\frac{1}{(1+e\cos(\sqrt{A}\,\varphi))^2}\ ,
\end{array}
\end{equation}
such that we obtain
\begin{equation}
\begin{array}{rcl}
u_{\mathrm{GR}.H^2}&=&\displaystyle\frac{C}{A}\,\frac{\alpha_{\mathrm{GR}}}{d}\left(\left(1+\frac{e^2}{2}\right)-\frac{e^2}{6}\,\cos(2\sqrt{A}\varphi)+\sqrt{A}\,e\,\varphi\sin(\sqrt{A}\,\varphi)\right)\ ,\\[6mm]
u_{H^2}(\varphi)&=&\displaystyle \frac{d^3\,H_0^2}{A\,J^2(1-e^2)}\left(\frac{\alpha_0\,GM}{c^2\,d}+\frac{(-4+e^2)+3e^2\cos(2\sqrt{A}\,\varphi)}{4(1-e^2)(1+e\,\cos(\sqrt{A}\,\varphi))}\right.\\[6mm]
&&\displaystyle\left. -\left(\frac{3}{2(1-e^2)}-\frac{\alpha_0\,GM}{c^2\,d}\right)\frac{2e\,\arctan\left(\sqrt{\frac{1-e}{1+e}}\,\tan\left(\frac{\sqrt{A}\,\varphi}{2}\right)\right)\sin(\sqrt{A}\,\varphi)}{\sqrt{1-e^2}}\right)\ ,
\end{array}
\label{u_GR_H2}
\end{equation}
where
\begin{equation}
\alpha_{GR}=\frac{3GM}{c^2\,d}\ .
\end{equation}
Both the solutions $u_{\mathrm{GR}.H^2}$ and $u_{H^2}$ have the same structure of the standard solution for Schwarzschild backgrounds~\cite{Kenyon,Gravitation}, the first term is a constant that can be neglected, the second term has a period that is a multiple of the period of solution $u_{0.H^2}$~(\ref{u_0_H}) contributing a small correction to the orbital period and the last term monotonically grows with increasing $\varphi$ contributing to the orbital perihelion advance. This last result is justified by noting that the analytic continuation of the inverse of a function corresponds to the argument of the function (in this way $\arctan(\tan \varphi)=\varphi$ increases monotonically with $\varphi$).
Due to the corrections to the Keplerian orbit's solution being small when compared to the dominant term, we can expand the trigonometric functions to lower order~\cite{Kenyon,Gravitation}:
\begin{equation}
\cos\left(\sqrt{A}\,\varphi\right) = 1-\frac{A^2\,\varphi^2}{2}+O\left(\varphi^4\right)\ ,
\end{equation}
\begin{equation}
\varphi\sin\left(\sqrt{A}\,\varphi\right)=2\frac{\sqrt{A}\,\varphi^2}{2}+O\left(\varphi^4\right)\ ,
\end{equation}
and
\begin{equation}
\arctan\left(\sqrt{\frac{1-e}{1+e}}\tan\left(\frac{\sqrt{A}\,\varphi}{2}\right)\right)=\frac{A\,\varphi^2}{2}\sqrt{\frac{1-e}{1+e}}+O\left(\varphi^4\right) \ .
\end{equation}
Hence, neglecting the constant terms in the solutions $u_{\mathrm{GR}.H^2}$ and $u_{H^2}$~(\ref{u_GR_H2}) and gathering the several terms and respective coefficients for these lower order expansions, we obtain the full solution $u$~(\ref{u_H2})
\begin{equation}
\begin{array}{rcl}
u&\approx&\displaystyle\frac{1}{d}\left(1+e\,\cos\left(\left(1-\frac{\Delta\varphi_{\mathrm{GR}}}{2\pi}-\frac{\Delta\varphi_{H^2}}{2\pi}\right)\,\varphi\right)\right)+u_{\mathrm{osc.GR}}+u_{\mathrm{osc}.H^2}\ ,\\[6mm]
\displaystyle\frac{\Delta\varphi_{\mathrm{GR}}}{2\pi}&=&\displaystyle\alpha_{\mathrm{GR}}\\[5mm]
\displaystyle\frac{\Delta\varphi_{H^2}}{2\pi}&=&\displaystyle-\frac{\delta_A}{2}+\alpha_{\mathrm{GR}}\,\delta_C\left(1+\frac{2e^2}{3d}\right)\\[5mm]
&&\hfill\displaystyle+\frac{d^3\,H_0^2}{(1-e)(1+e)^\frac{3}{2}}\,\left(\frac{\alpha_0}{c^2\,d}-\frac{3}{2(1-e^2)\,GM}\right)+O(H_0^4)\ ,\\[6mm]
u_{\mathrm{osc.GR}}&=&\displaystyle -\frac{\alpha_{GR}}{6d}\,e^2\cos(2\varphi)\ ,\\[6mm]
u_{\mathrm{osc}.H^2}&=&\displaystyle \frac{(\delta_C-\delta_A)\alpha_{\mathrm{GR}}}{6}\,e^2\cos(2\varphi)+(H_0\,d)^2\frac{-4+e^2+3e^2\cos(2\varphi)}{4(1-e^2)^2\,GM\,\left(1+e\cos\varphi\right)}+O(H_0^4)
\end{array}
\label{precession_H2}
\end{equation}
where $\Delta\varphi_{\mathrm{GR}}/(2\pi)$ is the standard perihelion advance per turn of the orbit due to General Relativity corrections on Schwarzschild backgrounds and $\Delta\varphi_{H^2}/(2\pi)$ is the perihelion advance per turn of the orbit due to the ELA metric background. As for the factor $u_{\mathrm{osc.GR}}$ it is the General Relativity oscillatory factor correction to the orbit solution obtained for Schwarzschild backgrounds and $u_{\mathrm{osc}.H^2}$ is the oscillatory factor correction due to the ELA metric background.

To compute the observable period correction to the orbits due to the ELA metric background it is enough to consider the definition of the constant of motion $J\,d\tau=-d\varphi/u^2$~(\ref{J}) and integrate the infinitesimal proper time displacement $d\tau$ over one turn of the orbit such that we obtain
\begin{equation}
T=-\frac{1}{J}\int_0^{2\pi}\frac{d\varphi}{u^2}\approx -\frac{1}{J}\int_0^{2\pi}d\varphi\,\frac{1}{u_0^2}\,\left(1-\frac{2u_{\mathrm{osc.GR}}}{u_0}-\frac{2u_{\mathrm{osc.H^2}}}{u_0}\right)\ .
\end{equation}
To directly compare the General Relativity corrections to the orbital period on the ELA metric background with the Keplerian orbital period and the General Relativity corrections on Schwarzschild backgrounds this integral can be factorized into the 3 components
\begin{equation}
\begin{array}{rcl}
T&=&T_0+\Delta T_{\mathrm{GR}}+\Delta T_{H^2}\ ,\\[6mm]
T_0&=&\displaystyle-\frac{1}{J_0}\int_0^{2\pi}d\varphi\frac{1}{u_0^2}\,=\, \frac{2\pi\,r_{\mathrm{orb}}^\frac{3}{2}}{\sqrt{GM}}\ ,\\[6mm]
\Delta T_{\mathrm{GR}}&=&\displaystyle+\frac{2}{J_0}\,\int_0^{2\pi}d\varphi\,\frac{u_{\mathrm{osc.GR}}}{u_0^3}\,=\,-\frac{3\pi\,\sqrt{GM}\,r_{\mathrm{orb}}^\frac{1}{2}\,e^4}{c^2(1-e^2)^2}\ ,\\[6mm]
\displaystyle\Delta T_{H^2}&=&\displaystyle-\frac{1}{2}\left(T_0+\Delta\,T_{\mathrm{GR}}\right)\left(\delta_A-\delta_B\right)-\frac{2}{|J|}\int_0^{2\pi}d\varphi\,\frac{u_{\mathrm{osc}.H^2}}{u_0^3}\\[4mm]
&\approx&\displaystyle -\left(\delta_A-\delta_B\right)\left(\frac{2\pi\,r_{\mathrm{orb}}^\frac{3}{2}}{\sqrt{GM}}-\frac{3\pi\,\sqrt{GM}\,r_{\mathrm{orb}}^\frac{1}{2}\,e^4}{2c^2(1-e^2)^2}\right)\\[4mm]
&&\displaystyle+\frac{3\pi\left(\delta_A-\delta_C\right)\,r_{\mathrm{orb}}^\frac{3}{2}\,e^4\,\sqrt{GM}}{c^2(1-e^2)}+\frac{\pi\,r_{\mathrm{orb}}^\frac{9}{2}\,(4+9e^2)\,H_0^2}{(GM)^\frac{3}{2}}+O(H_0^4)\ ,
\end{array}
\label{dT_H2}
\end{equation}
where $T_0$ is the classical Keplerian orbit period corresponding to the solution $u_0$~(\ref{u0.0}),
$\Delta T_{\mathrm{GR}}$ is the standard General Relativity period correction on Schwarzschild backgrounds
corresponding to solution $u_{\mathrm{osc.GR}}$~(\ref{precession_H2}) and $\Delta T_{H^2}$ is the General Relativity period correction on the ELA metric background corresponding to solution $u_{\mathrm{osc}.H^2}$~(\ref{precession_H2}).
\begin{table}[ht]
\begin{center}
{
\begin{tabular}{lcc}
\hline\noalign{\smallskip}
Planet&    $\displaystyle \Delta T_{\mathrm{GR}}\ (s/yr^{-1})$ & $\displaystyle \frac{\Delta \varphi_{\mathrm{GR}}}{2\pi}\ (arcsec/century^{-1})$\\
\noalign{\smallskip}\hline\noalign{\smallskip}
Mercury&  $+2.35\times 10^{-3}$   &$42.96$  \\[2mm]
Venus&    $+1.36\times 10^{-9}$   &$8.62$   \\[2mm]
Earth&    $+3.64\times 10^{-8}$   &$3.84$   \\[2mm]
Mars&     $+2.38\times 10^{-5}$   &$1.35$   \\[2mm]
Jupiter&  $+4.95\times 10^{-7}$   &$0.062$   \\[2mm]
Saturn&   $+4.24\times 10^{-7}$   &$0.014$   \\[2mm]
Uranus&   $+1.21\times 10^{-7}$   &$0.0024$   \\[2mm]
Neptune&  $+8.44\times 10^{-11}$  &$0.00077$   \\[2mm]
Pluto &   $+5.15\times 10^{-5}$   &$0.00042$   \\
\noalign{\smallskip}\hline
\end{tabular}}
\end{center}
\caption{Standard General Relativity corrections on the Solar Schwarzschild background with respect to the Solar Newtonian background (Keplerian orbits) to the orbital period $\Delta T_{\mathrm{GR}}$~(\ref{dT_H2})
	and orbital perihelion advance $\Delta \varphi_{\mathrm{GR}}/2\pi$~(\ref{precession_H2}) for each planet in the Solar System.}
\label{table.GR}
\end{table}
The values for the standard General Relativity corrections to the orbital period and orbital perihelion advance on Schwarzschild backgrounds are listed in table~\ref{table.GR}. These corrections correspond to orbital motion on Schwarzschild backgrounds being well known~\cite{Kenyon} and are already accounted for in the model employed in the numerical analysis of the Solar System dynamics which employs the PPN equations of motion. We recall that these corrections did constitute one of the original experimental tests of General Relativity.

In addition, although on Schwarzschild backgrounds the orbital radius is not varying over time, on expanding backgrounds such as the ones described by the ELA metric, it is expected that the radius does vary as the background expands. The analytical solutions computed so far do not allow to estimate such variation for the orbital radius as we have approximated the ELA metric background by a static background with fixed Hubble rate $H_0=H(t_0)$~(\ref{A.L_orbit}). In the next section, from conservation of angular momentum, we estimate the orbital radius variation by considering approximately circular orbits.

\section{Circular Orbits Approximation: Time Varying Orbital Radius\label{sec.radius}}

In this section, with the objective of estimating the orbital radius variation on backgrounds described by the ELA metric, we are analyzing circular orbits on such backgrounds. In the non-relativistic velocity limit and for relatively small values of the radial coordinate ($r\ll l_H=c/H$) the radial acceleration is
\begin{equation}
\ddot{r}_1\approx\displaystyle -c^2\Gamma^1_{\ 00}\approx -\frac{GM}{r^2}+\frac{2(GM)^2}{c^2\,r^3}+F_{H^2}+O(r^2\,H^4)\ .
\label{F_Newton_mod}
\end{equation}
Here dotted quantities represent derivation with respect to the coordinate time $t$. In the right hand side of the equation~(\ref{F_Newton_mod}), the first term is the usual classical Newton gravitational acceleration, the second term is the standard General Relativity correction on Schwarzschild backgrounds and the third term is the General Relativity correction of order $H^2$ for backgrounds described by the ELA metric~(\ref{g_generic})
\begin{equation}
F_{H^2}=+r\left(1-\frac{2GM}{c^2\,r}\right)^{\alpha_0}\left(1-\frac{(1-\alpha_0)GM}{c^2\,r}-(1+q)\left(1-\frac{2GM}{c^2\,r}\right)^{\frac{1}{2}-\frac{\alpha_0}{2}}\right)H^2\ .
\label{F_Newton_mod_I}
\end{equation}

To derive the orbital velocity let us consider the constant of motion corresponding to conservation of angular momentum, $J=-r^2\,d\varphi/d\tau$. Particularizing to circular orbits for which the orbital velocity is constant, $v_{\mathrm{orb}}=\sqrt{-r\,\ddot{r}_1}$, and considering the usual definition of angular velocity $\dot{\varphi}=\omega=v_{\mathrm{orb}}/r$ we obtain the following definition for the angular momentum $J_{\mathrm{circ}}^2=\left.-\gamma^2\,r^3\,\ddot{r}_1\right|_{r=r_{\mathrm{orb}}}$ such that for an orbit of radius $r_{\mathrm{orb}}$ we obtain
\begin{equation}
J_{\mathrm{circ}}^2\approx GM\,r_{\mathrm{orb}}\,\left(1+\left(\frac{H\,r_{\mathrm{orb}}}{c}\right)^2\left(1-\frac{2GM}{c^2\,r_{\mathrm{orb}}}\right)^{\alpha_0}\right)-F_{H^2}\,r_{\mathrm{orb}}^3\ .
\label{J_circ}
\end{equation}
Here $\gamma=dt/d\tau$ is the relativistic factor for the ELA metric~(\ref{g_generic}). Specifically, in the limit of non-relativistic velocity $\dot{x}^\mu\ll c$, it is
\begin{equation}
\frac{d^2r}{d\tau^2}\approx\gamma^2\,\ddot{r}_1\approx\frac{\ddot{r}_1}{1-\frac{2GM}{c^2r}-\left(\frac{H\,r}{c}\right)^2\left(1-\frac{2GM}{c^2\,r}\right)^{\alpha_0}}\ .
\label{gamma_approx_generic}
\end{equation}
For circular orbits, the main effect obtained due to the corrections on the ELA metric background
correspond to a time varying radius. Such effect can be verified from conservation of angular momentum. To lowest order in time, $H$ is expressed as $H(t)\approx H_0-q_0\,H_0^2\,t$ such that assuming a non-varying Gravitational constant $\dot{G}=0$ and non-varying mass $\dot{M}=0$ we are left with the only possibility of a time-varying orbital radius $\dot{r}_{\mathrm{orb}}=0$. Hence differentiating equation~(\ref{J_circ}) and solving the equation $\dot{J}_{\mathrm{circ}}=0$ for $\dot{r}_{\mathrm{orb}}$ we obtain, to lowest order in $H_0$, the time dependence of the orbital radius
\begin{equation}
\begin{array}{rcl}
\displaystyle\frac{\dot{r}_{\mathrm{orb}}}{r_{\mathrm{orb}}}&\approx&\displaystyle \frac{2q_0\,(H_0\,r_{\mathrm{orb}})^3}{GM}\,\left(1-\frac{2GM}{c^2\,r_{\mathrm{orb}}}\right)^{\frac{\alpha_0}{2}+\frac{1}{2}}\times\\[6mm]
&&\displaystyle\hfill\times\left(1+q_0-\left(1-\frac{(2-\alpha_0)GM}{c^2\,r_{\mathrm{orb}}}\right)\left(1-\frac{2GM}{c^2\,r_{\mathrm{orb}}}\right)^{\frac{\alpha_0}{2}-\frac{1}{2}}\right)+O(H_0^5)\ .
\end{array}
\label{dr_orb_circ}
\end{equation}
As expected from cosmological expansion this expression increases with the orbital radius $r_{\mathrm{orb}}$ and decreases with the mass $M$. Consistently at very large radius ($r\sim l_H=c/H$) the gravitational potential is negligible ($1/r\sim 0$) such that pure cosmological expansion is asymptotically recovered and no stable orbits exist ($\dot{r}_1/r\sim 2q_0^2\,H^3r^3>0$).

\begin{figure}
\includegraphics[width=\textwidth]{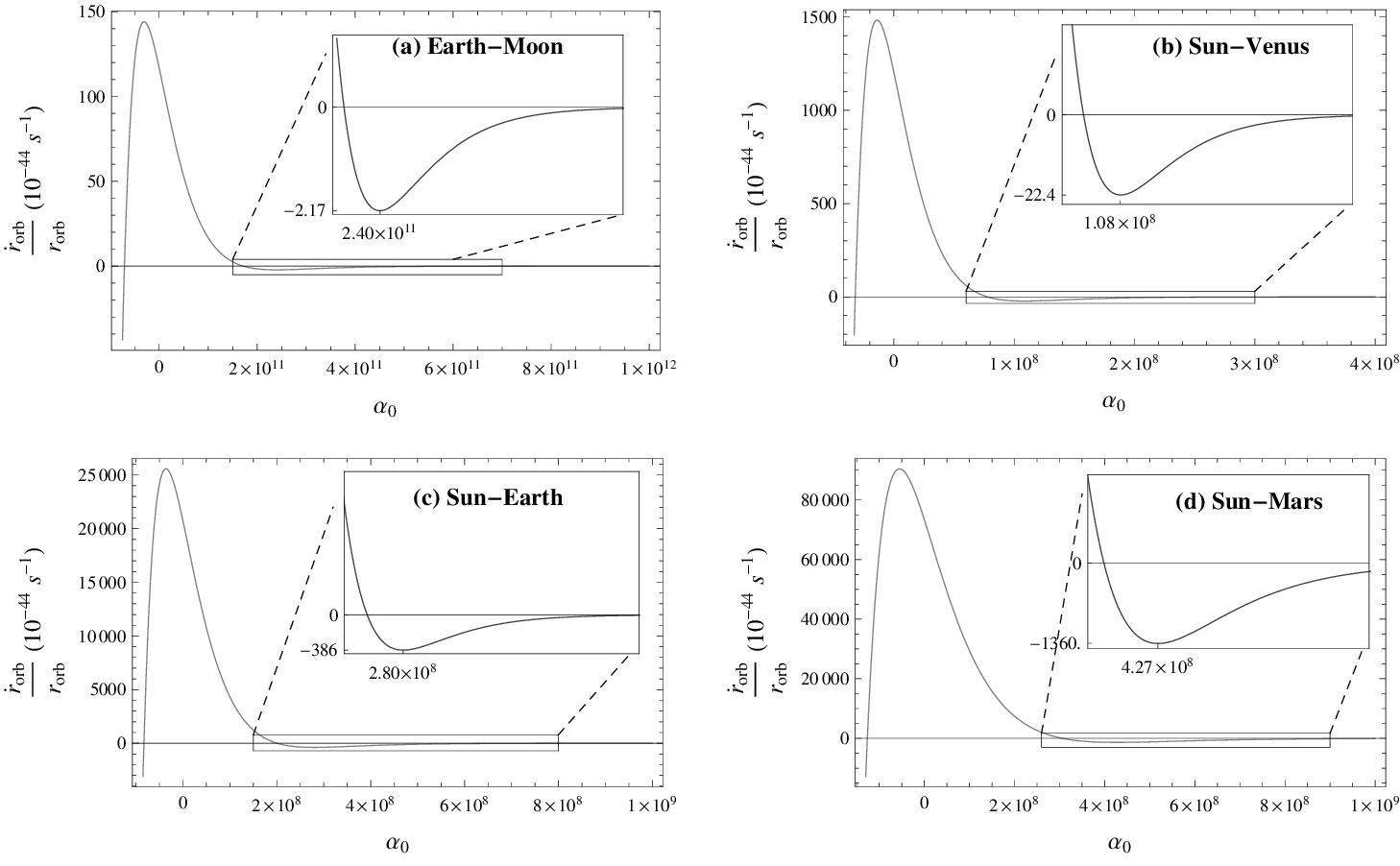}
\caption{Examples of the profiles of the time variation rate of the orbital
radius $\dot{r}_{\mathrm{orb}}/r_{\mathrm{orb}}$~(\ref{dr_orb_circ}) as a function of the
parameter $\alpha_0$ assuming the circular orbits approximation:\hfill\break
\ {\bf(a)} for the Earth-Moon orbit, the maximum positive variation is
$\dot{r}_{\mathrm{orb}}/r_{\mathrm{orb}}=1.440\times 10^{-42}\,s^{-1}$
corresponding to $\alpha_0=-3.084\times 10^{10}$ and the minimum negative variation for positive $\alpha_0$ is $\dot{r}_{\mathrm{orb}}/r_{\mathrm{orb}}=-2.173\times 10^{-44}\,s^{-1}$ corresponding
to $\alpha_0=2.396\times 10^{11}$ ;\hfill\break
\ {\bf(b)} for the Sun-Venus orbit, the maximum positive variation is
$\dot{r}_{\mathrm{orb}}/r_{\mathrm{orb}}=1.483\times 10^{-41}\,s^{-1}$
corresponding to $\alpha_0=-1.397\times 10^7$ and the minimum negative variation for positive $\alpha_0$ is $\dot{r}_{\mathrm{orb}}/r_{\mathrm{orb}}=-2.238\times 10^{-43}\,s^{-1}$ corresponding to $\alpha_0=1.085\times 10^8$ ;\hfill\break
\ {\bf(c)} for the Sun-Earth orbit, the maximum positive variation is
$\dot{r}_{\mathrm{orb}}/r_{\mathrm{orb}}=2.557\times 10^{-40}\,s^{-1}$
corresponding to $\alpha_0=-3.609\times 10^7$
and the minimum negative variation for positive $\alpha_0$ is $\dot{r}_{\mathrm{orb}}/r_{\mathrm{orb}}=-3.859\times 10^{-42}\,s^{-1}$ corresponding to $\alpha_0=2.803\times 10^8$ ;\hfill\break
\ {\bf(d)} for the Sun-Mars orbit, the maximum positive variation is
$\dot{r}_{\mathrm{orb}}/r_{\mathrm{orb}}=9.043\times 10^{-40}\,s^{-1}$
corresponding to $\alpha_0=-5.498\times 10^7$
and the minimum negative variation for positive $\alpha_0$ is $\dot{r}_{\mathrm{orb}}/r_{\mathrm{orb}}=-1.365\times 10^{-41}\,s^{-1}$ corresponding to $\alpha_0=4.271\times 10^8$.}
\label{fig.dr_alpha_circ}
\end{figure}

As for the specific dependence of the orbital radius variation on the parameter $\alpha_0$ it
is positive for small values of $\alpha_0\sim 0$ being of the same order of magnitude of the pure expansion effects, for growing positive values of this parameter, the radius variation decreases having a negative minimum value and then asymptotically vanishing in the limit $\alpha_0\to +\infty$ . This is actually expected, we note that in this limit the shift function is null, $\lim_{\alpha_0\to+\infty}(1-U_{\mathrm{SC}})=0$, such that we exactly recover the SC metric, hence a Ricci flat space-time for which $\dot{r}_1/r$ is exactly null for all orbits. As for growing negative values of this parameter the radius variation increases up to a maximum positive value and then decreases monotonically. For large negative values of this parameter $\alpha_0\ll 0$ the corrections with respect to Schwarzschild backgrounds become significantly higher with $\dot{r}_1/r<0$ being unbounded from below. As an example of the typical values of $\dot{r}_{orb}/r_{orb}$ as a function of the parameter $\alpha_0$ are plotted in figure~\ref{fig.dr_alpha_circ} the values of $\dot{r}_{orb}/r_{orb}$ for the Earth-Moon orbit and for Sun-Venus, Sun-Earth and Sun-Mars orbits.

We further note that the estimate for the orbital radius variation just computed is a valid approximation for elliptical orbits of small eccentricity $e\,\ll\,1$, hence a fairly good approximation for all planetary orbits in the Solar system except for Mercury's and Pluto's orbits for which $e\sim 0.2$ such that these estimates correspond, at most, to a rough approximation to the orbital radius variation for both these planets.

Next, considering the General Relativity corrections on backgrounds described by the ELA metric, we show that it is possible to map these corrections to the heuristic variation of the Gravitational constant by matching the corrections to Kepler's third law on such backgrounds to the fitted value of the parameter $\dot{G}$~(\ref{dAU_dG}).

\section{Modeling the Corrections to Kepler's Third Law\label{sec.AU}}

Next we will map the corrections to Kepler's third law obtained on the background described by the ELA metric~(\ref{g_generic}) directly to the heuristic fit to a variation of the Gravitational constant $\dot{G}\neq 0$~\cite{dG}. We recall that the original Solar System modeling is carried assuming the definition of the $AU$~(\ref{AU}) which is based on the classical Kepler's third law and, as discussed in the introduction, when this law is not considered to be a constraint~(\ref{AU}), both the values for the $AU$ and $G$ can be maintained fixed such that the measured deviations from the predictions of orbital motion on Schwarzschild backgrounds are accounted for by the ELA metric background. Aiming at mapping these effects to the heuristic fit of the variation of the Gravitational constant $\dot{G}/G$~(\ref{dAU_dG}) to planetary ephemerides obtained in~\cite{dG} we will discuss both the range measurement and the orbital motion predicted corrections on the backgrounds described by the ELA metric with respect to Schwarzschild backgrounds.

Specifically, for each planetary orbit, the effects that contribute to the heuristic variation of the Gravitational constant when considering Kepler's third law as a constraint~(\ref{AU}) are, the decrease of the Sun's mass, the orbital period corrections and the orbital radius variation. Hence, for a planetary orbit of average heliocentric semi-major radius $r_{orb}$ we obtain respectively the three distinct contributions
\begin{equation}
\left.\frac{\dot{G}}{G}\right|_{\mathrm{orb}}=\left.\frac{\dot{G}}{G}\right|_{\mathrm{orb}.\dot{M}_\odot}+\left.\frac{\dot{G}}{G}\right|_{\mathrm{orb}.\Delta T}+\left.\frac{\dot{G}}{G}\right|_{\mathrm{orb}.\dot{r}}\ ,
\label{dG_orbit}
\end{equation}
where the several contributions are
\begin{equation}
\begin{array}{lcl}
\displaystyle\left.\frac{\dot{G}}{G}\right|_{\mathrm{orb}.\dot{M}_\odot}&=&\displaystyle-\frac{\dot{M}_\odot}{M_\odot}\ (yr^{-1})\ ,\\[5mm]
\displaystyle\left.\frac{\dot{G}}{G}\right|_{\mathrm{orb}.\Delta T}&=&\displaystyle-2\,\frac{r_{\mathrm{orb}}^3}{GM_\odot}\left(\frac{2\pi}{T}\right)^2\frac{\Delta T_{H^2}}{T}\frac{T_{yr}}{T}\ (yr^{-1})\ ,\\[5mm]
\displaystyle\left.\frac{\dot{G}}{G}\right|_{\mathrm{orb}.\dot{r}}&=&\displaystyle+3\,\frac{r_{\mathrm{orb}}^3}{GM_\odot}\left(\frac{2\pi}{T}\right)^2\frac{\dot{r}_{\mathrm{orb}}}{r_{\mathrm{orb}}}\frac{T_{yr}}{T}\ (yr^{-1})\ .
\end{array}
\label{dG_3}
\end{equation}
In these expressions all quantities are expressed in IS units and the reference year is considered to be the Julian year define in IS seconds as
\begin{equation}
T_{yr}=31557600\ s\ .
\end{equation}

The contribution due to the variation of the Sun's mass is constant~(\ref{dAU_dM}) while both the corrections to the orbital period and the time variation of the orbital radius depend on the value of the metric parameter $\alpha_0$. To obtain a precise estimate for the values of these contributions it would be required a numerical analysis of the Solar System dynamics. To further proceed analytically let us consider the approximate analytical estimates discussed in the previous sections such that the expression for $\Delta T_{H^2}$ is given in~(\ref{dT_H2}) and the expression for $\dot{r}_{\mathrm{orb}}/r_{\mathrm{orb}}$ is given in~(\ref{dr_orb_circ}) corresponding to the corrections to the orbital period and orbital radius variation due to the expanding background described by the ELA metric with respect to the respective quantities computed on Schwarzschild backgrounds. These estimates are a fairly good approximation except for the orbital radius variation for the planet Mercury and the planet Pluto due to the relatively high eccentricity of their orbits. Nevertheless, as we are going to show next, the main contributions that are mapped to the variation of the Gravitational constant are due to the decrease of the Sun's mass~(\ref{dAU_dM}) and to the General Relativity corrections to the orbital period on backgrounds described by the ELA metric such that the contribution due to the orbital radius variation is negligible being lower than the remaining contributions by a factor of $10^{-27}$ at Mercury's orbit and by a factor of $10^{-24}$ at Pluto's orbit, hence being negligible for the estimate obtained. For completeness of the analytical expressions we are keeping this contribution in the following derivations.

\begin{figure}
\includegraphics[width=\textwidth]{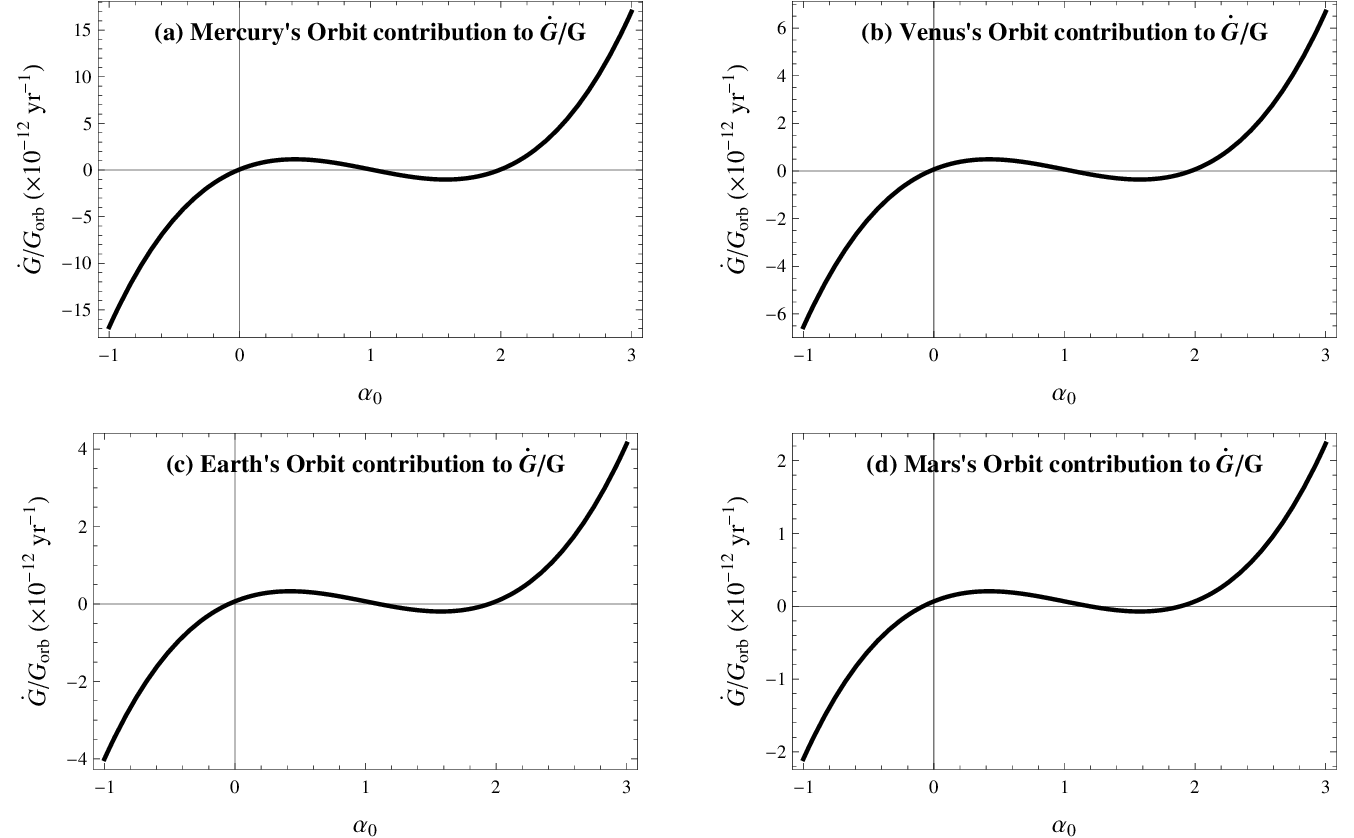}
\caption{Variation of the Gravitational constant $\left.\dot{G}/G\right|_{\mathrm{orb}}$~(\ref{dG_orbit}) as a function of the constant metric parameter $\alpha_0$ when Kepler's third law is considered as a constraint, for: {\bf(a)} the orbit of Mercury; {\bf(b)} the orbit of Venus; {\bf(c)} the orbit of Earth; {\bf(d)} the orbit of Mars.}
\label{fig.dG}
\end{figure}

For each planetary orbit these contributions have an inflection point near $\alpha_0=1$ such that for $\alpha_0\gg 0$ it is verified a significant increase of the variation of the Gravitational constant while for $\alpha_0\ll 0$ it is verified a significant decrease of the variation of the Gravitational constant. As an example, the dependence of the variation $\left.\dot{G}/G\right|_{\mathrm{orb}}$ on the values of the parameter $\alpha_0$ for the inner planets of the Solar System, Mercury, Venus, Earth and Mars is plotted in figure~\ref{fig.dG}.


As for the corrections to Kepler's third law as perceived for Earth based range measurements are computed by evaluating the difference between the corrections corresponding to the geodesic motion of Earth and the corrections corresponding to the geodesic motion of the planet for which the range measurement is being considered. Hence, for range measurements between Earth and any other planet in the Solar System, these corrections are mapped to a variation of the gravitational constant as
\begin{equation}
\left.\frac{\dot{G}}{G}\right|_{\mathrm{range}}=\left.\frac{\dot{G}}{G}\right|_{\mathrm{orb.Earth}}-\left.\frac{\dot{G}}{G}\right|_{\mathrm{orb.planet}}\ ,
\label{dG_range}
\end{equation}
where $\left.\dot{G}/G\right|_{\mathrm{orb.Earth}}$ and $\left.\dot{G}/G\right|_{\mathrm{orb.planet}}$ correspond to $\left.\dot{G}/G\right|_{\mathrm{orb}}$~(\ref{dG_orbit}) evaluated for the orbit of Earth and the orbit of each planet, respectively.

Further noting that, when performing numerical analysis and numerical integration of the Solar System dynamics, the fitted variation of the Gravitational constant is approximately a linear effect~\cite{dG} being independent of each planet's mass, a estimate for the average value of $\left.\dot{G}/G\right|_{\mathrm{orb}}$~(\ref{dG_orbit}) can be obtained by a simple average of the contributions due to each planetary orbit
\begin{equation}
\left<\frac{\dot{G}}{G}\right>_{\mathrm{orbit}}\,=\,\frac{\displaystyle\sum^{9}_{i=1}\left.\frac{\dot{G}}{G}\right|_{\mathrm{orb}.i}}{\displaystyle 9}\ .
\label{average_dG_orbit}
\end{equation}
As for the average value for the contribution of $\left.\dot{G}/G\right|_{\mathrm{range}}$ to the experimental data measurement can be estimated by an average weighted by the number of events $N_{i}$ divided by the respective rms residuals $\sigma_{i}$
\begin{equation}
\left<\frac{\dot{G}}{G}\right>_{\mathrm{range}}\,=\,\frac{\displaystyle\sum^{9}_{i=1,i\neq 3}\,\left.\frac{\dot{G}}{G}\right|_{\mathrm{range}.i}\,\frac{N_{i}}{\sigma_{i}}}{\displaystyle\sum^{9}_{i=1,i\neq 3}\frac{N_{i}}{\sigma_{i}}}\ .
\label{average_dG_range}
\end{equation}
In these expressions the index $i$ runs from $1$ to $9$ referring to the planets in the Solar System listed in table~\ref{table.planet_data} and $\left.\dot{G}/G\right|_{\mathrm{orb}.i}$ and $\left.\dot{G}/G\right|_{\mathrm{range}.i}$ are the variations of the Gravitational constant evaluated for each planetary orbit in the Solar System for heliocentric orbital motion~(\ref{dG_orbit}) and Earth based range measurements~(\ref{dG_range}), respectively. The values of the weights $w_{i}=N_{i}/\sigma_{i}$ are computed from table~2 and table~3 of~\cite{dG} being $\omega_1=4.65$, $\omega_2=35692.38$, $\omega_3=0$, $\omega_4=469279.08$, $\omega_5=74.86$, $\omega_6=348.02$, $\omega_7=62.40$, $\omega_8=64.66$ and $\omega_9=38.79$.

Hence, when considering the ELA metric background, there will be two distinct corrections which can be mapped into the heuristic fit to the variation of the Gravitational constant $\dot{G}/G$. Let us recall that the numerical analysis of the Solar System dynamics has, generally, two distinct procedures~\cite{Pitjeva,Fienga}. First the ephemerides are built considering as the base model only the well established General Relativity gravitation interactions on Schwarzschild backgrounds. Then the ephemerides are numerically integrated by considering a wide number of parameters which generally may include corrections to the gravitational interactions on Schwarzschild backgrounds, for example the PPN parameters $\gamma$, $\beta$ and $\alpha$, as well as the variation of the Gravitational constant, corresponding to the parameter $\dot{G}/G$. It is from this second numerical analysis that an unmodeled fit to the parameter $\left<\dot{G}/G\right>_{\mathrm{fit}}$ is obtained~(\ref{dAU_dG}). Therefore when mapping the ELA metric background gravitational interactions corrections to deviations from the gravitational interactions on Schwarzschild backgrounds it is required to consider corrections to both of these procedures. In particular when mapping the modeled correction to an heuristic variation of the Gravitational constant we obtain that the average value with respect to the ephemerides data is $\left<G\right>_{\mathrm{eph}}=G_0+\left<\dot{G}\right>_{\mathrm{range}}$ and, when fitting the ephemerides data to a heliocentric model of the Solar System, we obtain $\left<G\right>_{\mathrm{eph}}=G_0+\left<\dot{G}\right>_{\mathrm{fit}}-\left<\dot{G}\right>_{\mathrm{orb}}$, where the contribution $\left<\dot{G}\right>_{\mathrm{fit}}$ corresponds to the heuristic fit~(\ref{dAU_dG}). Hence matching these two expressions we obtain the following map between the fit to the parameter $\dot{G}/G$~(\ref{dAU_dG}) and the modeled average contributions~(\ref{average_dG_orbit}) and~(\ref{average_dG_range})
\begin{equation}
\displaystyle\left<\frac{\dot{G}}{G}\right>_{\mathrm{fit}}=\left<\frac{\dot{G}}{G}\right>_{\mathrm{orb}}+\left<\frac{\dot{G}}{G}\right>_{\mathrm{range}}\ .
\label{dG_fit_estimative}
\end{equation}

Generally we could consider a variation of the functional parameter $\alpha$ across the Solar System such that, for each planetary orbit, this metric parameter would be given by an approximately constant value $\alpha_{0.i}$. However a more accurate fit to ephemerides
requires a full numerical analysis including the gravitational corrections due to the ELA metric background. For analytical analysis purposes, let us simply consider an approximately constant coefficient $\alpha_0$ across the Solar System. Hence, from the map~(\ref{dG_fit_estimative}) and for the value of the fit~(\ref{dAU_dG}) we obtain
\begin{equation}
\left<\frac{\dot{G}}{G}\right>_{\mathrm{fit}}=1.65\pm 5.85\times 10^{-14}\ (yr^{-1})\ \ \Leftrightarrow\ \ \left.\alpha_0\right|_{H^2}=1.06^{+0.96}_{-1.14}\ .
\label{alpha_fit}
\end{equation}
We note that the dependence of the variation of the Gravitational constant on $\alpha_0$ has an inflexion point near $\alpha_0=1$ (see figure~\ref{fig.dG}), the relatively large uncertainty on the value of $\alpha_0$ is mainly due to the proximity to this inflexion point.
\begin{table}[ht]
\begin{center}
{\small
\begin{tabular}{lcccccc}
\hline\noalign{\smallskip}
Planet&    $\displaystyle \Delta\left.\dot{G}/G\right|_{\mathrm{orb}.\Delta T_{H^2}}$& $\displaystyle \Delta\left.\dot{G}/G\right|_{\mathrm{orb}.\dot{r}}$ & $\displaystyle \left.\dot{G}/G\right|_{\mathrm{orb}}$ & $\left.\dot{G}/G\right|_{\mathrm{range}}$\\
 &    $(yr^{-1})$ & $(yr^{-1})$ & $(yr^{-1})$ & $(yr^{-1})$\\
\noalign{\smallskip}\hline\noalign{\smallskip}
Mercury&  $-1.72^{+3.05}_{-3.05}\times 10^{-13}$  &$1.46\times 10^{-40}$   &$-1.05^{+3.36}_{-3.36}\times 10^{-13}$   &$+1.31^{+2.31}_{-2.31}\times 10^{-14}$\\[2mm]
Venus&    $-6.73^{+1.19}_{-1.19}\times 10^{-13}$  &$3.74\times 10^{-40}$   &$-3.22^{+1.50}_{-1.50}\times 10^{-13}$   &$+2.59^{+4.59}_{-4.59}\times 10^{-14}$\\[2mm]
Earth&    $-4.14^{+7.34}_{-7.34}\times 10^{-14}$  &$6.08\times 10^{-40}$   &$+2.56^{+1.04}_{-1.04}\times 10^{-13}$   &--                     \\[2mm]
Mars&     $-2.20^{+3.90}_{-3.90}\times 10^{-14}$  &$1.14\times 10^{-39}$   &$+4.50^{+7.00}_{-7.00}\times 10^{-14}$   &$-1.94^{+3.44}_{-3.44}\times 10^{-14}$\\[2mm]
Jupiter&  $-3.49^{+6.18}_{-6.18}\times 10^{-15}$  &$7.21\times 10^{-39}$   &$+6.35^{+3.72}_{-3.72}\times 10^{-14}$   &$-3.79^{+7.72}_{-7.72}\times 10^{-14}$\\[2mm]
Saturn&   $-1.41^{+2.49}_{-2.49}\times 10^{-15}$  &$1.79\times 10^{-38}$   &$+6.56^{+3.35}_{-3.35}\times 10^{-14}$   &$-4.00^{+7.09}_{-7.09}\times 10^{-14}$\\[2mm]
Uranus&   $-4.93^{+8.73}_{-8.73}\times 10^{-16}$  &$5.00\times 10^{-38}$   &$+6.65^{+3.19}_{-3.19}\times 10^{-14}$   &$-4.09^{+7.25}_{-7.25}\times 10^{-14}$\\[2mm]
Neptune&  $-2.51^{+4.45}_{-4.45}\times 10^{-16}$  &$1.00\times 10^{-37}$   &$+6.68^{+3.14}_{-3.14}\times 10^{-14}$   &$-4.12^{+7.30}_{-7.30}\times 10^{-14}$\\[2mm]
Pluto &   $-1.67^{+2.96}_{-2.96}\times 10^{-16}$  &$1.51\times 10^{-37}$   &$+6.68^{+3.13}_{-3.13}\times 10^{-14}$   &$-4.13^{+7.31}_{-7.31}\times 10^{-14}$\\
\noalign{\smallskip}\hline\noalign{\smallskip}
\end{tabular}}
\end{center}
\caption{Contributions to $\dot{G}/G$ mapped from the corrections to Kepler's third law on the ELA metric background with respect to Schwarzschild backgrounds. For each planet it is listed, in the first column the contribution due to the orbital period correction (the second term in equation~(\ref{dG_orbit})), in the second column the contribution due to the orbital radius variation (the third term in equation~(\ref{dG_orbit})) for which the uncertainty is at least 8 orders of magnitude below the quoted values, in the third column the total contribution to the variation of the Gravitational constant for heliocentric orbital motion $\left.\dot{G}/G\right|_{\mathrm orb}$~(\ref{dG_orbit}) and in the fourth column the total contribution to the variation of the Gravitational constant for Earth based range measurements $\left.\dot{G}/G\right|_{\mathrm{range}}$~(\ref{dG_range}).}
\label{table.planet_dG_H2}
\end{table}
In table~\ref{table.planet_dG_H2} are listed the values of the contributions from each planetary orbit in the Solar System to the estimate~(\ref{alpha_fit}). The correction to the orbital period per each Julian year in the range of the parameter $\alpha_0$ given in~(\ref{alpha_fit}) is approximately the same for all planetary orbits
\begin{equation}
\Delta T_{H^2}\,\frac{T_{yr}}{T}=-0.065\pm 1.16\times 10^{-6}\ s\,yr^{-1}\ ,
\end{equation}
where $\Delta T_{H^2}$ corresponds to the period correction per revolution for each planet~(\ref{dT_H2}). Although this correction is enough to map the heuristic fit to the variation of the $AU$, we note that it is negligible for most of other purposes, even for archaeological fits to the variation of the Solar System parameters we obtain at most a variation of the Earth year by $\pm 1.7\,h$ over a period of $10^9$ years, hence within the uncertainty of such estimate~\cite{Uzan}.

As for the values for the corrections to the orbital perihelion advance and orbital radii variation for each planet are listed in table~\ref{table.planet_df_dT}.
\begin{table}[ht]
\begin{center}
{
\begin{tabular}{lcc}
\hline\noalign{\smallskip}
Planet    &$\frac{\Delta \varphi_{H^2}}{2\pi}\,(mas\,century^{-1})$ & $\frac{\dot{r}_{\mathrm{orb}}}{r_{\mathrm{orb}}}\,(century^{-1})$ \\
\noalign{\smallskip}\hline\noalign{\smallskip}
Mercury              &$-5.36\times 10^{-12}$             &$1.17\times 10^{-39}$             \\[2mm]
Venus                &$-1.56\times 10^{-11}$             &$7.67\times 10^{-39}$             \\[2mm]
Earth                &$-2.53\times 10^{-11}$             &$2.03\times 10^{-38}$             \\[2mm]
Mars                 &$-4.54\times 10^{-11}$             &$7.17\times 10^{-38}$             \\[2mm]
Jupiter              &$-2.95\times 10^{-10}$             &$2.86\times 10^{-36}$             \\[2mm]
Saturn               &$-7.29\times 10^{-10}$             &$1.76\times 10^{-35}$             \\[2mm]
Uranus               &$-2.09\times 10^{-9}$             &$1.43\times 10^{-34}$             \\[2mm]
Neptune              &$-4.19\times 10^{-9}$             &$5.51\times 10^{-34}$             \\[2mm]
Pluto                &$-4.31\times 10^{-9}$             &$1.25\times 10^{-33}$             \\
\noalign{\smallskip}\hline
\end{tabular}}
\end{center}
\caption{Corrections to perihelion advance and orbital radius variation due to the corrections for the background described by the ELA metric with respect to the predictions on Schwarzschild backgrounds. The estimate uncertainty for each of these values is more than 8 orders of magnitude below the quoted values.}
\label{table.planet_df_dT}
\end{table}
These corrections are lower by more than 8 orders of magnitude when compared to the respective predictions for Schwarzschild backgrounds. The perihelion advance corrections are well below current measurement accuracy, the best fits attained have at most an accuracy of $0.1\, mas$~\cite{Fienga}. As for the orbital radius variation range from $\dot{r}_{\mathrm{orb}}/r_{\mathrm{orb}}\sim 10^{-32}\,century^{-1}$ for mercury up to $\dot{r}_{\mathrm{orb}}/r_{\mathrm{orb}}\sim 10^{-26}\,century^{-1}$ for Pluto, hence being well below any other estimate for these variations~\cite{Uzan}.

So far we have only discussed the corrections to Kepler's third law due to the background described by the ELA metric. In addition we recall that, generally, a variation of the Gravitational constant does not necessarily imply an orbital radius variation~\cite{dG,dG1,dG2} and further remark that the orbital radii variation obtained from numerical analysis of ephemerides in~\cite{dG} are due to the decrease of Sun's mass $\dot{M}_\odot$~(\ref{dAU_dM}), hence not directly comparable with the values listed in table~\ref{table.planet_df_dT}, for which the quoted orbital radius variation is due to the ELA metric background alone. Specifically, due to the decrease of the Sun's mass, for each planetary orbit it is verified a Newtonian variation of both the orbital periods~\cite{dG1,dG2} and the orbital radii~\cite{Oort_cloud,cl_corr}
\begin{equation}
\begin{array}{rcl}
\displaystyle\frac{\dot{T}_{\mathrm{orb}}}{T_{\mathrm{orb}}}&\approx&\displaystyle-\left(\frac{1}{2}+\frac{3}{2}\,\frac{1+e^2+2e\cos\,f}{1-e^2}\right)\,\frac{\dot{M}_\odot}{M_\odot}\ ,\\[5mm]
\displaystyle\frac{\dot{r}_{\mathrm{orb}}}{r_{\mathrm{orb}}}&\approx&\displaystyle-\frac{1+e^2+2e\cos\,f}{1-e^2}\,\frac{\dot{M}_\odot}{M_\odot}\ ,
\end{array}
\label{dr_dT_cl}
\end{equation}
where $f$ is the orbit's true anomaly given in table~\ref{table.planet_data}. The standard PPN equations of motion employed for the dynamical analysis of the Solar System do not include the effects due to the Sun's mass variation, hence these corrections must also be taken in consideration when mapping the fitted value of the Gravitational constant to the gravitational corrections due to the background described by the ELA metric. Specificaly, directly from Kepler's third law we obtain the following Newtonian corrections due to the orbital periods and radii variations~(\ref{dr_dT_cl})
\begin{equation}
\begin{array}{rcl}
\displaystyle\left.\frac{\dot{G}}{G}\right|_{\mathrm{Newton.}\dot{T}_{\mathrm{orb}}}&=&\displaystyle-2\frac{\dot{T}_{\mathrm{orb}}}{T_{\mathrm{orb}}}\ ,\\[5mm]
\displaystyle\left.\frac{\dot{G}}{G}\right|_{\mathrm{Newton.}\dot{r}_{\mathrm{orb}}}&=&\displaystyle+3\frac{\dot{r}_{\mathrm{orb}}}{r_{\mathrm{orb}}}\,\frac{r_{\mathrm{orb}}^3}{GM_\odot}\,\left(\frac{2\pi}{T_{\mathrm{orb}}}\right)^2\ .
\end{array}
\end{equation}
We note that both these contributions to the corrections to Kepler's third law have opposite signs such that, considering the background described by the ELA metric, the main contribution to the variation of the Gravitational constant $\dot{G}/G$ is still due to the orbital period corrections $\Delta T_{H^2}$. Including the corrections~(\ref{dr_dT_cl}) on the previous analisys we obtain the final estimate for the metric parameter $\alpha_0$
\begin{equation}
\left<\frac{\dot{G}}{G}\right>_{\mathrm{fit}}=1.65\pm 5.85\times 10^{-14}\ (yr^{-1})\ \ \Leftrightarrow\ \ \alpha_0=0.98^{+1.06}_{-1.01}\ .
\label{alpha_fit_1}
\end{equation}
Hence it is not excluded the interpretation that our result simply corresponds to a statistical flutuation of the experimental measurements. Also we remark that the analysis carried here was computed by considering the average effect of two body interactions, specifically the Sun and each of the planets. A full numerical analysis of orbital ephemerides is required to properly include the many body interactions in the Solar System and eventualy obtain a more accurate estimate for the metric parameter.

With respect to the ELA metric background we note that the value of the metric parameter $\alpha_0$ parameterizes the local anisotropic corrections with respect to the isotropic cosmological background, specifically space-time is locally isotropic for $\alpha_0=0$ which corresponds to the isotropic background described by the McVittie metric~\cite{McVittie}. The value of $\alpha_0\approx 1.06$~(\ref{alpha_fit}) corresponds to a relatively small perturbation to the isotropic background which, as has been shown, corresponds to relatively small corrections to the orbital parameters. Consistently with this discussion we remark that, when compared with a isotropic variation of the Gravitational constant $\dot{G}/G$, the corrections due to the ELA metric background to the orbital period are relatively more relevant than the corrections to the orbital perihelion advance and orbital radius variation. It is due to the background anisotropy between the radial direction and angular directions that such effect is attainable. In addition we recall that, when considering point-like massive objects, as we approach the SC horizon the metric exponent $\alpha$ should be greater or equal to $\alpha(r_{\mathrm{SC}})=3$ to ensure that space-time is singularity free at this horizon. This requirement is not absolutely necessary as the real Sun is not a point-like mass being instead an extended spheroid, hence without an event horizon. Nevertheless we further note that the uncertainty on the estimate of the constant $\alpha_0$ is relatively large and we may expect that a functional parameter $\alpha$ varying across the Solar System would allow for a better fit to planetary ephemerides, hence we may conjecture that its value should be decreasing with growing heliocentric distances being close to $\alpha=3$ near the Sun. This discussion is not conclusive being required a numerical analysis of the Solar System dynamics including the corrections due to the ELA metric background to actually verify if such a profile for the values of $\alpha$ is the best fit to planetary motion in the Solar System.

\section{Conclusions\label{sec.conclusions}}

In this work we have mapped the corrections to Kepler's third law on backgrounds described by the ELA metric~(\ref{g_generic}) to the heuristic variation of the Gravitational constant $\dot{G}/G$ estimated from numerical analysis of the Solar System dynamics on Schwarzschild backgrounds. These corrections plus the decrease of the Sun's mass by radiation emissions fully account for the fitted value of the variation of the Gravitational constant~(\ref{dAU_dG}). Reflecting the anisotropic nature of the ELA metric background, the more relevant contribution to such modeling is due to the orbital period corrections, being the contributions to the orbital radii variation negligible.

The constant value for the metric parameter that matches the quoted variation of the Gravitational constant is $\alpha_0=0.98^{+1.06}_{-1.00}$~(\ref{alpha_fit_1}), hence relatively close to the value $\alpha_0=0$ which corresponds to the isotropic background described by the McVittie metric. For completeness let us further note that other effects which may be relevant on backgrounds described by the ELA metric such as the corrections to the Doppler shift for range measurements, are negligible for this value of the metric parameter, being of the same order of magnitude of the effects attributed to the isotropic cosmological expansion~\cite{McV_light,Pioneer_ELA}. Following the same arguments we further conclude that, within the Solar System the contribution to the cosmological mass-energy density within the Solar System due to the ELA metric background is negligible~\cite{DM_orbits} (for further details see~\cite{Pioneer_ELA}).

Hence we have shown that the heuristic variation of the Gravitational constant $\dot{G}/{G}$~\cite{AU_1,dG} can alternatively be mapped and modeled by the ELA metric background parameterizing the corrections to gravitational interactions within the Solar System without considering Kepler's third law as a constraint through the definition of the $AU$~(\ref{AU}). Such construction allows for a fixed constant value both for the $AU$ and the Gravitational constant $G$ independently of the original definition of the $AU$ as has recently been suggested~\cite{dG,Fienga}. For analytical analysis purposes we have considered a constant metric parameter $\alpha_0$ for which we obtain a relatively high uncertainty. More generally a radially symmetric functional parameter with varying value across the Solar System would allow for a significant reduction of such uncertainty as well as allowing to match the distinct unmodeled estimates for $\dot{G}$ for the several planets in the Solar System~\cite{Uzan,G}. The results obtained here are enough to motivate a numerical analysis of the Solar System dynamics including the ELA metric background corrections to gravitational interactions. This framework solves the problem of the unwelcome variation of the measurement projection standard (whether the $AU$, whether the Gravitational constant $G$) and constitutes as well a playground for testing the ELA metric background in the most well known of all the astrophysical systems, the Solar System. We leave such study to another work.

\ \\ {\bf \large Acknowledgements}\\
This work was supported by grant SFRH/BPD/34566/2007 from FCT-MCTES.
Work developed in the scope of the strategical project of GFM-UL PEst-OE/MAT/UI0208/2011.



\end{document}